\documentclass[preprint,10pt]{aastex6}
\usepackage{graphicx}
\usepackage{amsmath}
\newcommand{\kgsins}[1]{{#1}}
\newcommand{\kgsdel}[1]{}
\newcommand{\tksins}[1]{{#1}}
\newcommand{\cbins}[1]{{#1}}

\begin{document}

\title{The rotation period distributions of 4--10 Myr T Tauri stars in Orion OB1: 
New constraints on pre-main-sequence angular momentum evolution}
\author{Md Tanveer Karim\altaffilmark{1}, Keivan G.\ Stassun\altaffilmark{2,11}, C\'esar Brice{\~n}o\altaffilmark{3}, A.\ Katherina Vivas\altaffilmark{3}, Stefanie Raetz\altaffilmark{4}, 
Cecilia Mateu\altaffilmark{5}, Juan Jos\'e Downes\altaffilmark{5}, 
Nuria Calvet\altaffilmark{6}, Jes\'us Hern\'andez\altaffilmark{5}, 
Ralph Neuh\"auser\altaffilmark{7}, Markus Mugrauer\altaffilmark{7}, Hidenori Takahashi\altaffilmark{8}, Kengo Tachihara\altaffilmark{8}, Rolf Chini\altaffilmark{9,12}, Gustavo A. Cruz-Dias\altaffilmark{10,13}, Alicia Aarnio\altaffilmark{6}, David J.\ James\altaffilmark{3}, Moritz Hackstein\altaffilmark{9}}

\altaffiltext{1}{University of Rochester,
Department of Physics and Astronomy,
500 Joseph C.\ Wilson Blvd,
Rochester, NY 14627 USA}

\altaffiltext{2}{Vanderbilt University, Department of Physics \& Astronomy,
VU Station B 1807, Nashville, TN 37235 USA}

\altaffiltext{3}{Cerro Tololo Interamerican Observatory, Casilla Postal 603, La Serena 1700000, Chile}

\altaffiltext{4}{Scientific Support Office, Directorate of Science and Robotic Exploration, European Space Research and Technology Centre
(ESA/ESTEC), Keplerlaan 1, 2201 AZ Noordwijk, The Netherlands}

\altaffiltext{5}{Centro de Investigaciones de Astronom{\'\i}a (CIDA), Apartado Postal 264, M\'erida 5101-A, Venezuela}

\altaffiltext{6}{University of Michigan, Astronomy Department, 311 West Hall 1085 South University Ave, Ann Arbor, MI 48109 USA}

\altaffiltext{7}{Astrophysikalisches Institut und Universit\"ats-Sternwarte, FSU Jena, Schillerg\"a\ss chen 2-3, 07745 Jena, Germany}

\altaffiltext{8}{Gunma Astronomical Observatory, 6860-86 Nakayama, Takayama-mura, Agatsuma-gun, Gunma 377-0702 Japan}

\altaffiltext{9}{Astronomisches Institut, Ruhr-Universit\"at Bochum, Universit\"atsstr. 150, D-44801 Bochum, Germany}

\altaffiltext{10}{NASA - Ames Research Center, Moffet Field, CA 94035, USA}


\altaffiltext{12}{Instituto de Astronom{\'i}a, Universidad Católica del Norte, Avenida Angamos 0610, Antofagasta, Chile}

\altaffiltext{13}{BAER Institute, USA}


\begin{abstract}
Most existing studies of the angular momentum evolution of young stellar populations have focused on the youngest ($\lesssim$1--3 Myr) T Tauri stars \kgsdel{which are usually still associated with their natal gas and hence easier to identify}. In contrast, the angular momentum distributions of older T Tauri stars ($\sim$4--10 Myr) have been less studied, even though they hold key insight to understanding stellar angular momentum evolution 
at a time when protoplanetary disks have largely dissipated and when models therefore predict changes in the rotational evolution that can in principle be tested. We present a study of photometric variability among 1,974 
confirmed T Tauri members of various sub-regions of the 
Orion OB1 association, and with ages spanning 4--10 Myr,
\kgsdel{We use} \kgsins{using} optical time-series from three different surveys 
\kgsdel{the Centro de Investigaciones de Astronom{\'\i}a (CIDA) Variability Survey of Orion (CVSO), the Young Exoplanet Transit Initiative (YETI), and a Kitt
Peak National Observatory (KPNO) campaign}. 
For 564 of the stars ($\sim$32\% of the weak-lined T Tauri stars and $\sim$13\% of the classical T Tauri stars in our sample) we detect statistically significant periodic variations
which we attribute to the stellar rotation periods,
\kgsins{making this one of the largest samples of T Tauri star rotation periods yet published}. 
\kgsdel{We do not find strong differences in the rotation period distributions for weak-lined versus classical T Tauri stars in our 4--10 Myr sample.}
We \kgsdel{do, however,} observe a clear change in the overall rotation period distributions over the age range 4--10 Myr, with the progressively older sub-populations exhibiting systematically faster rotation.
This result is consistent with angular momentum evolution model predictions of an important qualitative change in the stellar rotation periods starting at $\sim$5 Myr,
an age range for which very few observational constraints were previously available.
\end{abstract}

\section{Introduction}



\kgsins{As pre--main-sequence (PMS) stars contract toward the main sequence, they would be expected to rapidly spin up to near breakup velocity as a consequence of angular momentum conservation. However, numerous surveys of rotation periods among low-mass PMS stars clearly show that these stars typically rotate at a small fraction of breakup, despite significant contraction in stellar radius \citep[e.g.,][]{Stauffer:1987}. Consequently,}
one of the long-standing questions in star-formation research is 
\kgsins{to determine} the physical processes that govern 
\kgsins{the angular momentum evolution of low-mass PMS stars}
\kgsins{\citep[for a recent comprehensive review, see, e.g.,][]{Bouvier:2014}}. 

Central to this work is the direct measurement of rotation periods for large samples of T Tauri stars (TTSs), ideally over a large range of ages spanning key transition points in the evolution of PMS stars and the protoplanetary disks that they harbor. In particular, some models of PMS angular momentum evolution invoke magnetic interactions between a TTS and its circumstellar disk in order to drain stellar angular momentum during the first few Myr of the star's contraction toward the main sequence, for example through a ``disk locking" mechanism \citep[e.g.,][]{Ostriker1995}, an accretion powered stellar wind \citep[e.g.,][]{Matt2012}, and/or extreme coronal mass ejections \citep[e.g.,][]{Aarnio2013}.

Because PMS disks are known to dissipate on a timescale of a few Myr \citep[e.g.,][]{Haisch2000}, these models predict a qualitative change in the stellar rotational evolution immediately following the accretion wind/disk dominated phase, such that the stellar rotation should become faster as the stars continue to contract toward the main sequence but now lack an efficient braking mechanism.
Therefore, measurements of TTS rotation periods at ages when disks have very recently been dissipated---i.e., at $\sim$5--10 Myr---are essential for testing these model predictions.

A number of studies of TTS rotation periods have been conducted over the past few decades, resulting in measured rotation periods for thousands of TTSs. 
Rotation periods are usually measured from periodic modulations of the TTS light curve, which may arise from cool starspots and/or hot accretion spots \citep[e.g.,][]{herbst94}. 
Other phenomena such as disk occultations \citep[e.g.,][]{rodriguez15} can cause quasi-periodic signals, but these usually occur on very different timescales and do not show the sinusoid-like character of spot-modulated rotation signals. 

The most extensively monitored regions include 
\citep[see, e.g.,][]{Irwin2008,Henderson2012} 
the Taurus–Auriga association (age $\sim$1--3 Myr), the Orion Nebula Cluster (ONC; $\sim$1--2 Myr), the Lagoon Nebula (NGC 6530; $\sim$1--2 Myr), NGC 2264 ($\sim$3 Myr), NGC 2362 ($\sim$3--4 Myr), and IC 348 ($\sim$4--5 Myr), \kgsins{where the quoted nominal cluster ages and age ranges are from the compilation in \citet{Bouvier:2014}}. 
However, 
the rotational properties and evolution of older TTSs, and specifically at ages of $\sim$5--10 Myr relevant to testing angular momentum evolution models, remain poorly sampled by existing studies. 
\cbins{This is in part because of the lack of significant samples of young stars at this age range. 
Nearby sparse stellar aggregates like the $\sim 10$ Myr old TW Hya, or the $\sim 5-8$ Myr $\epsilon$ and $\eta$ Cha groups contain only about 20 stars each \citep[e.g.,][2015]{luhman2004,lyo2004,murphy2012}.
}

\cbins{In order to address the scarcity of older TTS, \cite{briceno01,briceno05}, and more recently Brice\~no et al. 2016 (in preparation, hereafter B16), 
have carried out a large scale survey encompassing $\sim 180$ square degrees across the Orion OB1 association, 
the Centro de Investigaciones de Astronom{\'\i}a Variability Survey of Orion (CVSO),
the with the goal of identifying and characterizing the older pre-main sequence stars across this extended star-forming complex.
Until now, it had been challenging to survey TTSs over this entire region because of the very large 
area spanned by the Orion OB1 sub-associations on the sky (see Fig.~\ref{fig:sd}).

\begin{figure}[ht]
	\centering
	\includegraphics[width=0.7\textwidth,clip,trim=0 125 0 150]{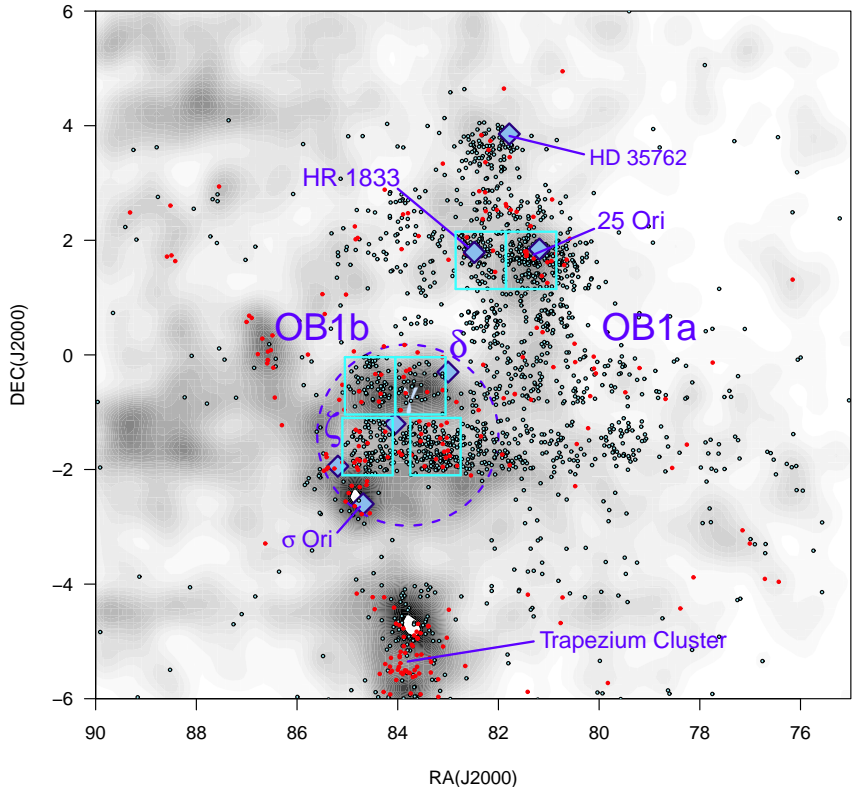}
	\caption{Spatial distribution of the 1,974 confirmed TTSs discussed in this paper. The background grayscale map represents the surface density of photometric TTS candidates in the CVSO (B16), computed using a two-dimensional kernel density estimator \citep{venables02,Rmanual}. CTTS are plotted as red dots and WTTS as small circles. The main Orion OB1 sub-associations are labeled in blue. Following \cite{briceno05}, the OB1b region is considered to span the area within the dashed circle. The belt stars as well as other bright B2 stars are plotted with large diamond symbols. The 25 Ori cluster and the newly discovered HR 1833 and HD 35762 stellar aggregates are indicated. The cyan squares show the KPNO 0.9m MOSAIC fields. The YETI field on 25 Ori is encompassed within the KPNO field centered on that cluster. The CVSO data span all the area in the figure.}
	\centering
	\label{fig:sd}
\end{figure}

Here we use the CVSO sample of nearly 2000 TTS to look for periodic photometric variability and measure rotation periods 
in the various sub-regions of the Orion OB1 star-forming complex. 
\kgsins{Orion OB1
is ideal for this study because its sub-associations comprise TTSs over the full age range of interest \citep{briceno05},
from $\sim$4 Myr (OB1b) 
to $\sim$10 Myr (OB1a)}. 
We use photometric monitoring data from several surveys of Orion OB1: the long time-baseline CVSO, the Young Exoplanet Transit Initiative \citep[YETI][]{neuhauser11}, and a dedicated high-cadence survey from the Kitt Peak National Observatory (KPNO) newly reported here.
}
We report rotation periods for 564 TTSs (out of 1,974 surveyed), examine how the rotation period distributions depend on age, and compare the distributions to the predictions of PMS angular evolution models.

In Sec.~\ref{section:data} we describe our data sources and data processing. In Sec.~\ref{section:method}, we describe the methods we used to measure rotation periods and their statistical significance. We present the resulting rotation period distributions 
in Sec.~\ref{section:results} and discuss the implications for
predictions of rotational evolution models in Sec.~\ref{sec:disc}.
Finally, we conclude with a brief summary in Sec.~\ref{section:conclusion}.

\section{Observations and Light Curve Production}
\label{section:data}

The multi-epoch optical photometric $V$, $R_C$, and $I_C$-band time-series for the 1,974 TTS members of Orion OB1 discussed here were collected from three different surveys, and incorporated into a single dataset for each star in each photometric bandpass.  \cbins{First, the CVSO \citep[][B16]{briceno01,briceno05}} which constitutes the source of the entire sample and provides the longest temporal baseline observations. Then, the Young Exoplanet Transit Initiative \citep[YETI;][]{neuhauser11} observations of the 25 Ori cluster, and the NOAO project 2005B-0529 observations of selected fields in Orion OB1, build on the CVSO by adding many more epochs in different bands, and with much higher temporal sampling, especially the YETI data. Table~\ref{tab-cadence} summarizes the observing strategy and cadence
of each one of these surveys.

\begin{deluxetable}{cccc}
\tabletypesize{\footnotesize}
\tablecolumns{4}
\tablewidth{0pt}
\tablecaption{Details of the observational strategy in each survey and photometric band}
\tablehead{\colhead{} & \colhead{QUEST} & \colhead{KPNO} & \colhead{YETI}}
\startdata
\cutinhead{V band}
N of stars                   & 1823        &             &               \\
Range in mag                 & 12.6 - 19.8 &             &               \\
Dates                        & 1999-2011   &             &               \\
Minimum Cadence (minutes)    & 2           &             &               \\
Time Baseline (years)     & 12.0        &             &               \\
N$_{min}$, N$_{max}$, N$_{median}$ & 1, 179, 54  &             &               \\
\cutinhead{R$_C$ band}
N of stars                   & 1795        &             & 287            \\
Range in mag                 & 12.5 - 19.8 &             & 12.6 - 18.0    \\
Dates                        & 1998-2011   &             & 2010-2012      \\
Minimum Cadence (minutes)    & 2           &             & 0.03           \\
Time Baseline (years)     & 12.0        &             & 2.1            \\
N$_{min}$, N$_{max}$, N$_{median}$ & 1, 202, 55  &             & 1, 13155, 2098 \\
\cutinhead{I$_C$ band}
N of stars                   & 1804        & 528         &                \\
Range in mag                 & 12.1 - 18.8 & 12.2 - 16.9 &                \\
Dates                        & 1998-2011   & 2006        &                \\
Minimum Cadence (minutes)    & 2           & 11          &                \\
Time Baseline (years)     & 12.1        & 0.02        &                \\
N$_{min}$, N$_{max}$, N$_{median}$ & 1, 184, 44  & 22, 35, 30  &                \\
\enddata
\label{tab-cadence}
\end{deluxetable}

Each of the surveys are described in Sections \ref{sec:cvso}--\ref{sec:kpno}, and the final production of the light curves for analysis is described in Section \ref{section:finallc}.

\subsection{CVSO}\label{sec:cvso}
The CVSO was carried out at the Llano del Hato National Astronomical Observatory in Venezuela, with the QUEST CCD mosaic camera ($8000 \times 8000$ pixels) on the 1-m (clear aperture) Schmidt telescope, with a plate scale of 1.02" pixel$^{-1}$ and field of view of 5.4 deg$^{2}$ \citep{baltay02}. This $V$, $R_C$, and $I_C$-band multi-epoch survey, covering $\sim 180$ deg$^2$ of the Orion OB1 association, spans a time baseline of 12 years, from December 1998 to February 2011. \kgsins{The survey was done in drift-scan mode with different filters covering each row of 4 CCDs in the array, providing quasi-simultaneous photometry in 3 or 4 bands\footnote{A row of CCDs was non-functional during several observing seasons.}. Since the observations were made having several different science cases in mind, different filter combinations (for example, BVRI, UBUV, VRI, VVI) were used along the years. This resulted in a very unhomogeneous set of observations in each photometric band. The exposure times are set in drift-scan mode to the time it takes to a star to cross over a single CCD, which for the QUEST camera is 140~s at the celestial equator.}

The CVSO was conceived to identify candidate TTSs based on their colors and photometric variability, using color-magnitude diagrams. Spanning apparent magnitudes $V\approx$13--19.5, it includes stars in the range $0.7 > M/M_{\odot} > 0.2$, with a mean mass of $\sim$0.5 M$_\odot$ as estimated via the PMS evolutionary tracks of \citet{baraffe98}. 
The survey was optimized to find and map the low-extinction, young off-cloud populations of the OB association which have been largely neglected in most studies of Orion OB1. 

The first years of data were used by \citet{briceno01} and \citet{briceno05} to detect candidate TTSs across Orion OB1 (see Fig.~\ref{fig:sd}). They confirmed $\sim$ 252 members through follow up spectroscopy, including the discovery of new stellar aggregates like the 25 Ori cluster (see next section). Subsequent observations
increased both the area coverage and the temporal sampling. The sample of 1,974 TTSs discussed here, includes those initial 252 young stars, 66 members of the 25 Ori cluster first identified in \cite{downes2014a}, and an additional 1,656 members, also confirmed through spectroscopy, drawn from the extended study of B16. 
The survey observations log, data reduction and calibration of the photometry up to the year 2008 are described in \citet{briceno05} and \citet{mateu12}. Here we added an additional $\sim 200$ observations taken during the period 2008-2011. 
All the new observations were handled in a similar way as the previous data and were normalized to the Mateu et al's catalog by determining the
average zero point difference among the non-variable stars. 

Based on their coordinates, we extracted the full time-series of the 1974 TTSs from the
resulting catalog.
Each star in the CVSO sample has a median of 54 observations in $V$, 55 in $R_C$, and 44 in the $I_C$ band, but some stars can have up to a maximum 
179, 202 and 184 epochs in $V$, $R_C$, and $I_C$ respectively. The spatial distribution of the CVSO TTS population is shown in Figure \ref{fig:sd}.

\subsection{YETI}\label{sec:yeti}
The Young Exoplanet Transit Initiative (YETI) is an international collaboration created with the goal of finding transiting exoplanets in nearby young stellar clusters. YETI is formed by a network of small telescopes (0.2 to 2.6m aperture) spread worldwide across a large range in longitude, which enables almost continuous observations of targets over several days \citep{neuhauser11}. One of the first clusters targeted by YETI was the 25 Ori cluster in the Orion OB1a sub-association, discovered by the CVSO \citep{briceno05,briceno07}. It contains $\sim 200$ TTSs within 1 degree, surrounding the early-B type stars 25 Orionis. All YETI observations were taken in the $R_C$ band.

The 25 Ori cluster was observed by the 0.6/0.9-m Schmidt type telescope at Jena Observatory (Germany), the two 5.9" telescopes at Observatorio Cerro Armazones (OCA, Chile) and the 1.5m reflector at the Gunma  Astronomical Observatory in Japan, over four observing campaigns during the years 2010--2013. 



The Jena Schmidt-type telescope was equipped with the optical Schmidt Telescope Camera \cite[STK][]{mugrauer10}, with an e2v 42-10 $2048 \times 2048$ detector, yielding a plate scale of 1.55" pixel$^{-1}$ and a field of view of $53' \times 53'$, thus encompassing most of the cluster. The Jena 50s exposures, all taken through the $R$ filter, were centered on 25 Ori. A total of 8506 individual exposures were obtained in 108 nights. 
The CCD images were bias and dark corrected, and then flatfielded, using the {\sl CCDPROC} package in IRAF\footnote{IRAF is distributed by the National Optical Astronomy Observatory, which is operated by the Association of Universities for Research in Astronomy (AURA) under a cooperative agreement with the National Science Foundation.}.
The Gunma 1.5m reflector observations were carried out by obtaining 60~s integrations in $R$ with the Gunma Low-resolution Spectrograph and Imager (GLOWS), which has a e2v CCD55-30 $1250 \times 1152$ pixel detector with a 0.6" pixel$^{-1}$ scale, covering a field of $12.5' \times 11.5'$ field of view. Observations were obtained during 4 nights
in year 2010. 
The raw data were bias subtracted and flat-fielded using the {\sl CCDPROC} package in IRAF.
The Observatorio Cerro Armazones observations were done in the R-band using the RoBoTT (Robotic Bochum TWin Telescope), which consists of twin Takahashi 150mm aperture apochromatic astrographs, each equipped with an Apogee U16M camera with a KAF-16803 $4096 \times 4096$ pixel CCD, providing a $2.7^\circ \times 2.7^\circ$ field of view with 2.37" pixel$^{-1}$ scale. The 60~s exposures were centered on 25 Ori, spanning an area much larger than the cluster.
OCA data were obtained during all YETI seasons. 

The photometry from the reduced images coming from the various observatories was carried out in a consistent and uniform way. First, instrumental magnitudes were derived by doing simple aperture photometry with a dedicated IRAF task called {\sl chphot}. Accurate positions for every star were derived with {\sl Sextractor} \citep{bertin96} or {\sl Eclipse} Jitter \citep{devillard97}, and then sky positions were determined using {\sl astrometry.net} \citep{lang10}, so stars in images from different observatories could be correctly matched using {\sl Topcat} \citep{taylor05}.
Differential photometry was done using the {\sl Photometry} software developed by \cite{broeg05}, which takes a weighted average of all stars in a frame to create an artificial comparison star. The advantage of this method is that it directly produces light curves for all the stars referenced to this synthetic standard star. Each night was processed independently using this scheme, and following \cite{errmann14} the data from different nights was combined using the night-to-night differences of the of stars selected as constant standards.

Because of its nature, the YETI observations produced more densely sampled time-series than the the CVSO or the KPNO observations (see below), with the average star having a median of 2098 measurements, and some up to 13,155 measurements. Thanks to the wide field of view of the OCA observations, a total of 287 of the CVSO TTSs ended up being observed within the YETI campaigns, though the ones with the largest number of measurements are within the 
central $\sim 40'$ of the cluster. \kgsins{Given the size of the telescopes involved, only the brightest of the CVSO TTSs have YETI observations}

\subsection{KPNO 0.9m data}\label{sec:kpno}
During the nights of Jan 8--15, 2006, we used the 0.9m telescope with the $8000 \times 8000$ pixel MOSAIC imager at the Kitt Peak National Observatory (KPNO), Arizona, USA, to obtain $I_C$-band time series observations of several regions in the Orion OB1 association, including the 25 Ori cluster in the OB1a sub-association, and fields in the OB1b sub-association, under NOAO program 2005B-0529. 
The strategy was to carry out a high cadence photometric time-series study of the two main sub-associations in Orion OB1: the 10 Myr old OB1a region and the 4 Myr old OB1b sub-association. 
The KPNO 0.9m fields shown in Figure~\ref{fig:sd} encompass 528 of the CVSO TTSs, each star having a median of 30, and a maximum of 35, observations. 
The data were processed following the procedures described in \citet{stassun99}.

\subsection{Production of final light curves}
\label{section:finallc}

In order to compile the data from the three surveys described above we used the CVSO as the base reference. 
For each one of the
KPNO-MOSAIC observations, we identified suitable comparison stars which were located within $10\arcmin$ of each TTS, 
had photometric errors $<0.05$ mags in both the CVSO and KPNO data, and were flagged as non-variable in the CVSO catalog. 
For each TTS only comparison stars
observed with the same CCD in the MOSAIC images were used in order of assure that differences among the chips did not affect the
zero point calculations. A ($3\sigma$-clipping) zero-point difference was calculated with the selected comparison stars and applied to
the TTS data. Between $15-40$ comparison stars were used to calculate the zero-point difference for each TTS.

A similar procedure was done for the YETI data but since some of the telescopes involved in this project produce shallow images (see above) which
is translated in less matches with the CVSO catalog, we expanded the search radius of comparison stars to $30\arcmin$. With this
search radius, we used $\sim 100-400$ comparison stars for the JENA data and $\sim 40-80$ for the OCA and Gunma data. 

In summary (\ref{tab-cadence}, the $V$-band time series contain data for 1,823 TTS (92\% of the sample) from the CVSO exclusively. There are 1,795 TTS (91\%) 
with $R_C$-band time series. The $R_C$ observations come from CVSO and YETI, although only a subsample of the TTS were within the YETI area and
have exquisite time sampling. Finally, the $I_C$-band time-series exists for 1,810 stars (92\% of the sample) with observations from CVSO and KPNO.
The sub-sample observed with KPNO have a good sampling concentrated within a range of 6-days. Every one of the 1,974 TTS in the total sample has
time series data in at least one photometric band. \kgsins{The multiple observations obtained during single nights with the KPNO and YETI surveys were very useful to break possible 1-day aliases (more discussion on this below).}

\section{Methods for Periodicity Analysis}
\label{section:method}

\begin{figure}[!ht]
\centering
\includegraphics[width=.6\textwidth]{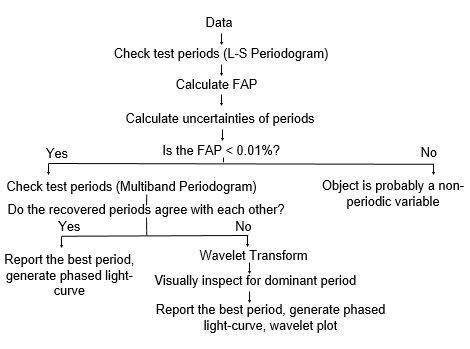}
\caption{Visual representation of the steps in the periodicity analysis methodology.}
\label{fig:fchart}
\end{figure}

We determined the variability amplitude and standard deviation for each star in each of the $V$, $R_C$, and $I_C$ filters by combining all data from the different surveys 
(see Section~\ref{section:finallc}). 
We also clipped each light curve with a 3$\sigma$ outlier rejection to reduce the effect of outliers. 
Next, we used several methods to determine the most probable period for each of the 1974 TTSs. We applied the Generalized Lomb-Scargle Periodogram \citep[GLS;][]{zechmeister09} and the Multiband Periodogram \citep{vanderplas15} to calculate the probable periods, \tksins{the uncertainty formula described by \citet{kovacs81} to calculate the uncertainties in periods}, and the Baluev method \citep{baluev08} to calculate the False Alarm Probability (FAP). We also used the Wavelet transform method \citep{bravo14} to determine the probable periods for stars with conflicting periods. All of these methods were combined into a single pipeline, shown in Figure~\ref{fig:fchart}, that provides the most likely period for each time-series while reducing the effect of bias. 

The following summarizes the procedure, with the various steps represented in Figure~\ref{fig:fchart} and described more fully in the subsections that follow:
\begin{enumerate}
\item In order to avoid problems caused by sparse sampling, we search for periodicity in the light curve of a given star in a given filter if it contains at least 30 measurements.
\item The time series in each filter 
is passed through the Generalized Lomb-Scargle Periodogram. The outputs are the maximum periodogram power and the period corresponding to that peak. \tksins{Furthermore, the uncertainties in the periods are calculated using the formula derived by \citet{kovacs81}.} 
\item Aliased periods are removed, i.e., periods that are within the range $(x - 0.05)$~d and $(x + 0.05)$~d, where $x$ is any integer between 1 and 30 d.
\item The maximum periodogram power and the number of data points in the time series are used as input to calculate the False Alarm Probability according to the method described by \citet{baluev08}. Stars that have FAP $>$ 0.01\%
in all three bands are considered non-periodic. We adopted this conservative FAP threshold so that, in our full sample of 1974 stars, we expect fewer than 1 false positive.
\kgsins{The subsequent steps are therefore not intended to further limit false positives but rather to ensure that the periods produced from different light curves and different period-finding methods are consistent.}
\item Time series are passed through the Multiband Periodogram and only one period is obtained as output for each star. 
\kgsins{This ensures that, if there is a period, we are not selecting an alias of the real period.}
\item Aliased periods from the Multiband Periodogram Analysis are removed. 
\kgsins{This ensures that there is not disagreement among the real periods identified in the multiple passbands of a given star.}
\item Periods from the Multiband Periodogram Analysis are compared to the GLS periods. If the star has only one GLS period with the associated FAP $<$ 0.01\%, 
then that period is reported as the best period. If the star has more than one GLS period, then we check to see if at least one of the GLS periods agree with the Multiband period. We consider two periods to be in agreement if they are within 0.5~d of one another. 
\kgsins{This ensures that periods obtained from the different methods agree to within a reasonable tolerance (below we will examine the period distributions with a binning resolution of 1~d).}
\item If none of the GLS periods agree with the Multiband period, or if no Multiband period was obtained, we then used the wavelet transform method to identify a dominant period. If a dominant period is seen, we report that value as the best period. If we see no dominant period, no period is reported.
\kgsins{Note that this final step is found to be necessary for only 10 stars in the study sample (Sec.\ \ref{section:wavelet}).}
\end{enumerate}

In the following subsections, we discuss the above steps in detail.
Appendix~\ref{section:applc} contains the phased light curves of all of the periodic stars. 
Appendix~\ref{section:appwt} contains the wavelet transform plots of ten stars whose periods were determined with the help of the Wavelet Analysis method, discussed in detail in Section~\ref{section:wavelet}.
Appendix~\ref{section:apptable} contains a table with the final periods
and FAPs of all the stars deemed to be periodic.

\subsection{Periodogram Analysis}
\label{section:gls}

We used VARTOOLS \citep{hartman08} to calculate the Generalized Lomb-Scargle (GLS) periodogram \citep{zechmeister09} of the sample stars, in order to determine their rotation periods. The GLS periodogram is well suited to our data for period searching because, unlike Fast Fourier Transform, it does not require an evenly spaced time-series and it is weighed by points, not time intervals. 
The VARTOOLS code is based on the algorithm outlined in \citet{press92}. The period search range was chosen to be 0.1--30~d, with the short period cutoff dictated by the fastest cadence of our light curves (mainly the high-cadence YETI and KPNO data; see above) and the long period cutoff 
\kgsins{based on previous work that finds 30~d to be among the very longest TTS periods \citep[e.g.,][and references therein]{Bouvier:2014}}. 
\tksins{The sampling was done in frequency space with a sampling of 1/$\Delta T$, where $\Delta T$ is the time-span of the observation for a given star in a given band.}
We used the GLS periodogram in each of the filters separately; thus, we obtained up to three possible periods for each star. 

\subsubsection{False Alarm Probability}
\label{section:FAP}
We considered a peak to be significant if its False Alarm Probability (FAP) is less than 0.01\%. 
\citet{suveges15} discusses the relative merits of four methods to assess the significance of periodogram peaks. Of the four methods, the $F^{M}$ method \citep{czerny12} is perhaps the most widely used, through which the FAPs are determined using Monte-Carlo simulations. However, \citet{suveges15} argue that this method tends to underestimate the FAP value and thus yields more false positives. In contrast, two other methods---the Generalized Extreme-value Distribution (GEV) method \citep{suveges14} and the Baluev method \citep{baluev08}---are more conservative than the $F^{M}$ method when determining the value of the FAP. 
We opted to use the Baluev method because 
the FAP calculation is less computationally expensive for data such as ours, and we sought to apply one method consistently across the very large number of multi-band light curves that comprise our data. 

Considering only detected periods with FAPs less than 0.01\%, 
we additionally excluded periods that were likely to represent aliases arising from the diurnal sampling of the data. 
We therefore excluded periods within the range $(x - 0.05)$ and $(x + 0.05)$, where $x$ is any integer between 1--30~d, as has been done in previous work \citep[e.g.,][]{stassun99,Irwin2008,Henderson2012}. 
\kgsins{Note that in principle there may also be aliases at sidereal day spacings as distinct from diurnal spacings, an effect sometimes observed in ultra-high-precision asteroseismic data. However in practice the very small difference (0.9973~d versus 1.0000~d) is not discernible in the periodograms from our ground-based data.}

\subsubsection{\tksins{Period Uncertainty Calculation}}
We calculated the uncertainties of the periods based on the formula derived by \citet{kovacs81}: 
$\delta \nu = 3 \sigma / 4T \sqrt{N} A$,
where $\delta P = P^{2} \delta \nu$, 
$P$ is the calculated period, $\sigma$ is the uncertainty in the data, $T$ is the total time spanned by the data, $N$ is the number of data points and $A$ is the amplitude of the time series. For example, for the star CVSO 6, $P_{V}$ = 5.62 days, $\sigma_{V}$ = 0.066 mag, $T_{V}$ = 4336.1 days, $N_{V}$ = 65 and $A_{V}$ = 0.271 mag, which implies that $\delta P_{V}$ = 1.65$\times 10^{-4}$~d.  For all cases, the formal period uncertainties were found to be less than 1\%.

\subsubsection{Multiband Periodogram Analysis}
To take more full advantage of cases where we have light curve observations in multiple filters, we also used the Multiband Periodogram method of \citet{vanderplas15} for period searching. 
This method extends the Lomb-Scargle periodogram  \citep{lomb76,scargle82} 
by simultaneously
modelling the light curves in each of the filters as arbitrary truncated Fourier series, with the period and the phase shared across all filters. 
We considered this method, instead of only considering the GLS periodograms for all of the filters separately, because there are cases where the period detected in one filter is different from the period detected in another filter. One possible reason for such difference is that the measurements in various filters are not taken at precisely
the same time, and in some cases there may be large differences in cadence and/or number of measurements in the different filters. Thus, by using the Multiband Periodogram method as well as the GLS method above to search for periodicity in the data, we could check which period is more likely.  

A limitation of the Multiband Periodogram is that there is not a developed FAP statistic. 
Therefore, we utilized the Multiband Periodogram  
as a check on the periods recovered by the GLS method. If a star has a GLS period in only one band and passes the FAP and aliasing tests, we report the GLS period and classify that star to be periodic. 
If a star has more than one GLS period, i.e.\ periods in different bands, and passes the FAP and aliasing test, we check whether at least one of the GLS periods agree with the Multiband period. 
If there is a match, we classify that star to be periodic and report the agreed period. If none of the GLS periods match the Multiband period, the wavelet analysis method (Sec.~\ref{section:wavelet}) is used.

\subsection{Wavelet Analysis}
\label{section:wavelet}
We used the Morlet Wavelet Transform method\footnote{http://www.phy.uct.ac.za/courses/python/examples/moreexamples.html} \citep{bravo14}
to search for periods of certain stars. The wavelet transform method takes the time series as input and generates a two-dimensional period vs. time plot. This allowed us to see whether a certain star had one dominant period during the entire time baseline or if there were multiple periods dominating at different times. A star having more than one dominant period during the entire time baseline may imply that multiple processes are taking place in or near the vicinity of the star at the same time. For example, if a star has two dominant periods, one of the periods may be the rotation period of the star while the other period may be related to the inner accretion disk and/or magnetosphere.

We used this method to estimate the rotation period for the few stars where none of the periods recovered by the GLS method agree with the period recovered by the Multiband periodogram to within 0.5~d. 
By examining the wavelet output, we attempted to qualitatively determine which period appears most likely to be the dominant one. If we detect a dominant period for a given star, we report that period and classify that star as periodic, but flag it as tentative in the final list of reported periods. If we do not detect any dominant period, we classify it as non-periodic. We classified ten stars as periodic with the help of this method and report their probable periods with an asterisk (*) symbol in the Best Period column in Appendix~\ref{section:apptable}. Appendix~\ref{section:appwt} contains the wavelet transform plots of these ten stars.

\subsection{Period Recoverability}

In order to assess the sensitivity of our data and period search methods to recover periods for different types of stars in our sample, we performed a period recoverability test \kgsins{intended to represent the recoverability that is typical for our data}. 
\kgsins{Strictly speaking, the recoverability will be a function of multiple parameters, some of which we can characterize for the sample as a whole (e.g., photometric noise, amplitude of spot signal, period of spot signal), whereas others are specific to each star (e.g., specific number of measurements and sampling pattern). There may also be complexities due to the rotational spot signals not being strictly sinusoidal in all cases, such as if the star has not a single spot but a group of spots. As a compromise given our large sample of stars and data, we followed the approach of several previous authors \citep[e.g.,][]{Irwin2008,Henderson2012} and adopted a simple sinusoidal signal with a number of data points and a sampling pattern that are typical for our sample, while varying the period, the spot amplitude, and the noise over a large range of parameter space.} 

\kgsins{Specifically,} we generated 240,000 synthetic light curves with the same time sampling and noise properties as the real light curves \kgsins{of a representative star} in our sample, and injected into these synthetic light curves sinusoidal signals of various periods (0.1--30 d) and amplitudes (0.01--0.30 mag), which are typical of TTSs. We then applied our period search algorithms to these synthetic light curves as we did for the real data, and recorded the fraction of correctly recovered periods as a function of the stellar magnitude, period, and amplitude.

\begin{figure}
    \gridline{\fig{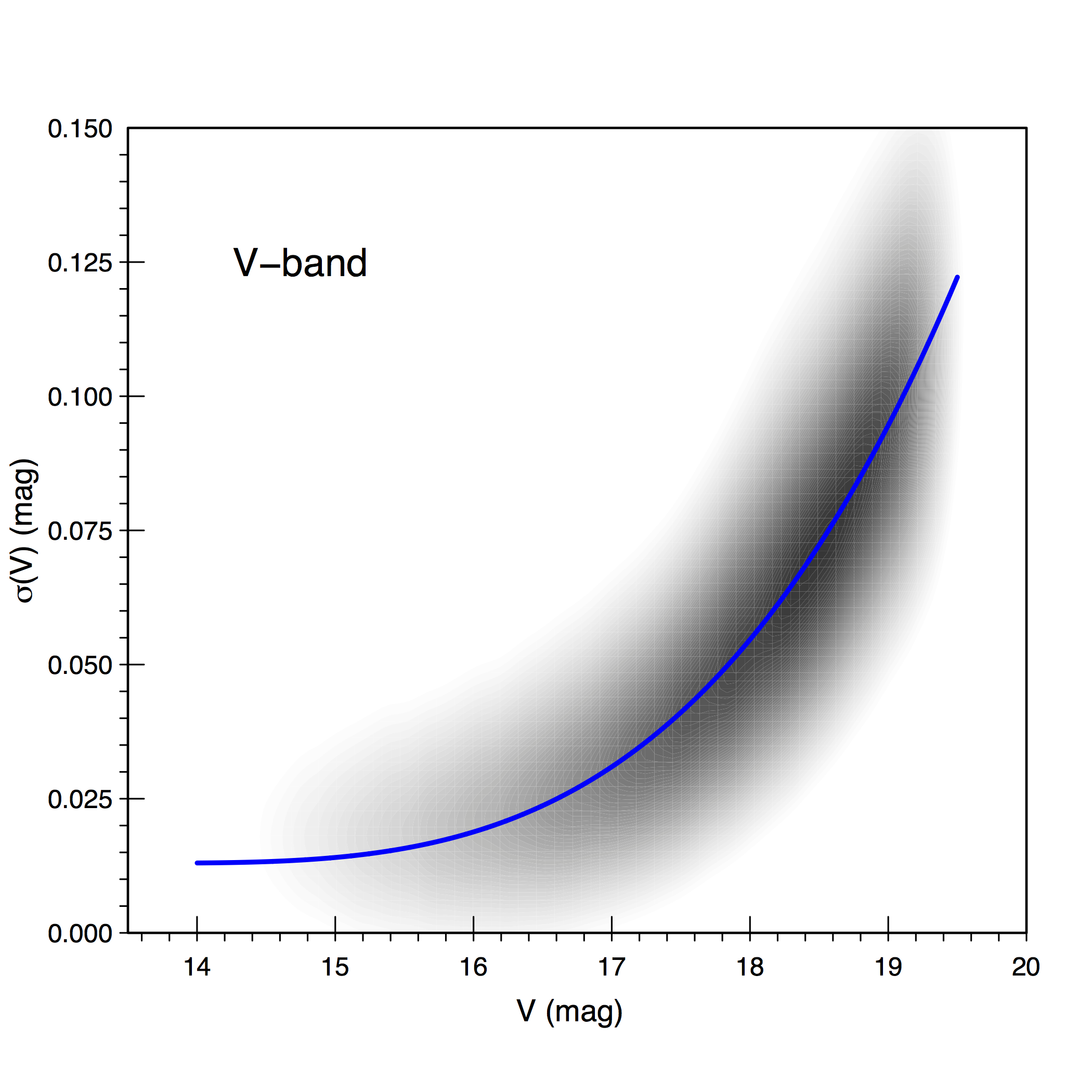}{0.31\textwidth}{(a)}
          \fig{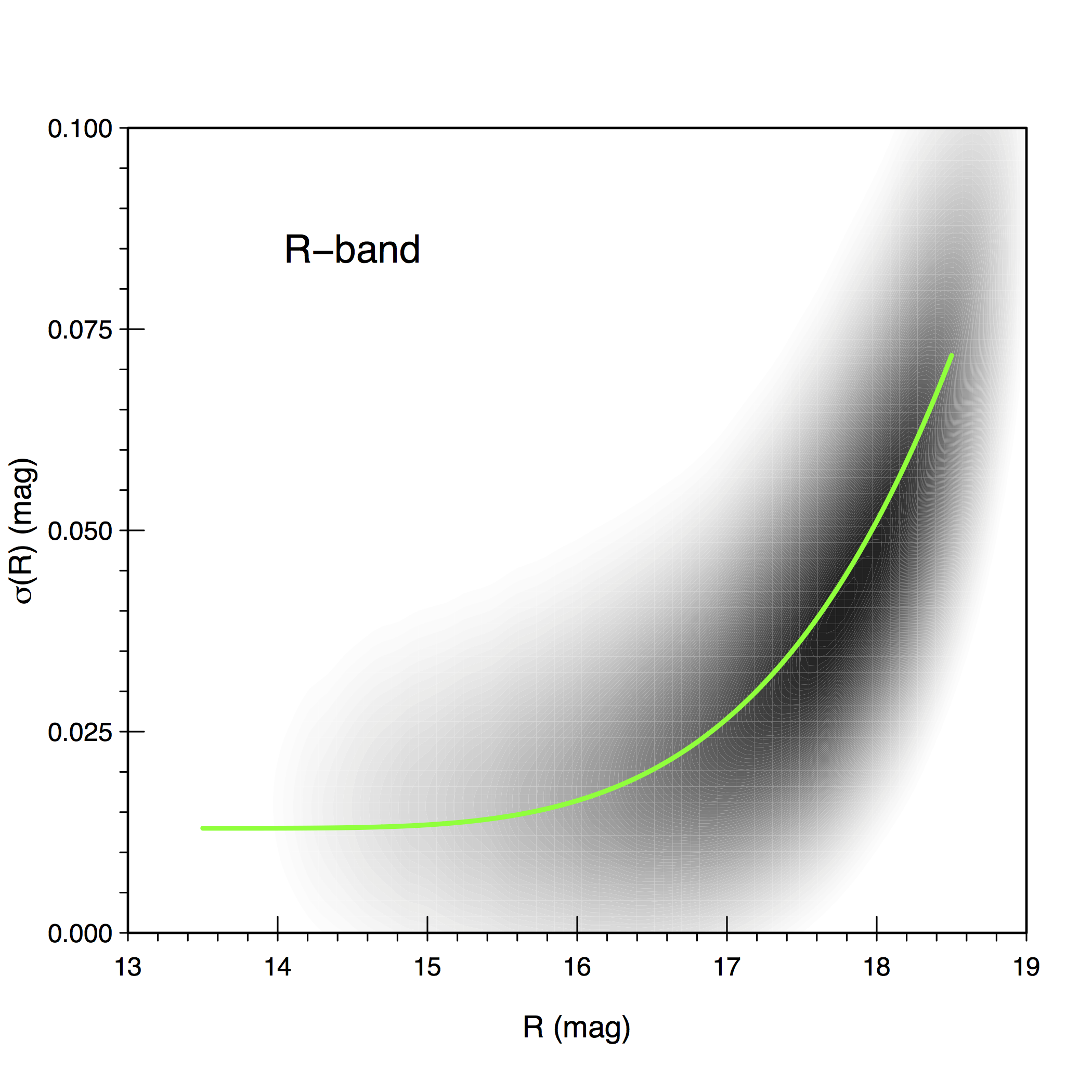}{0.31\textwidth}{(b)}
          \fig{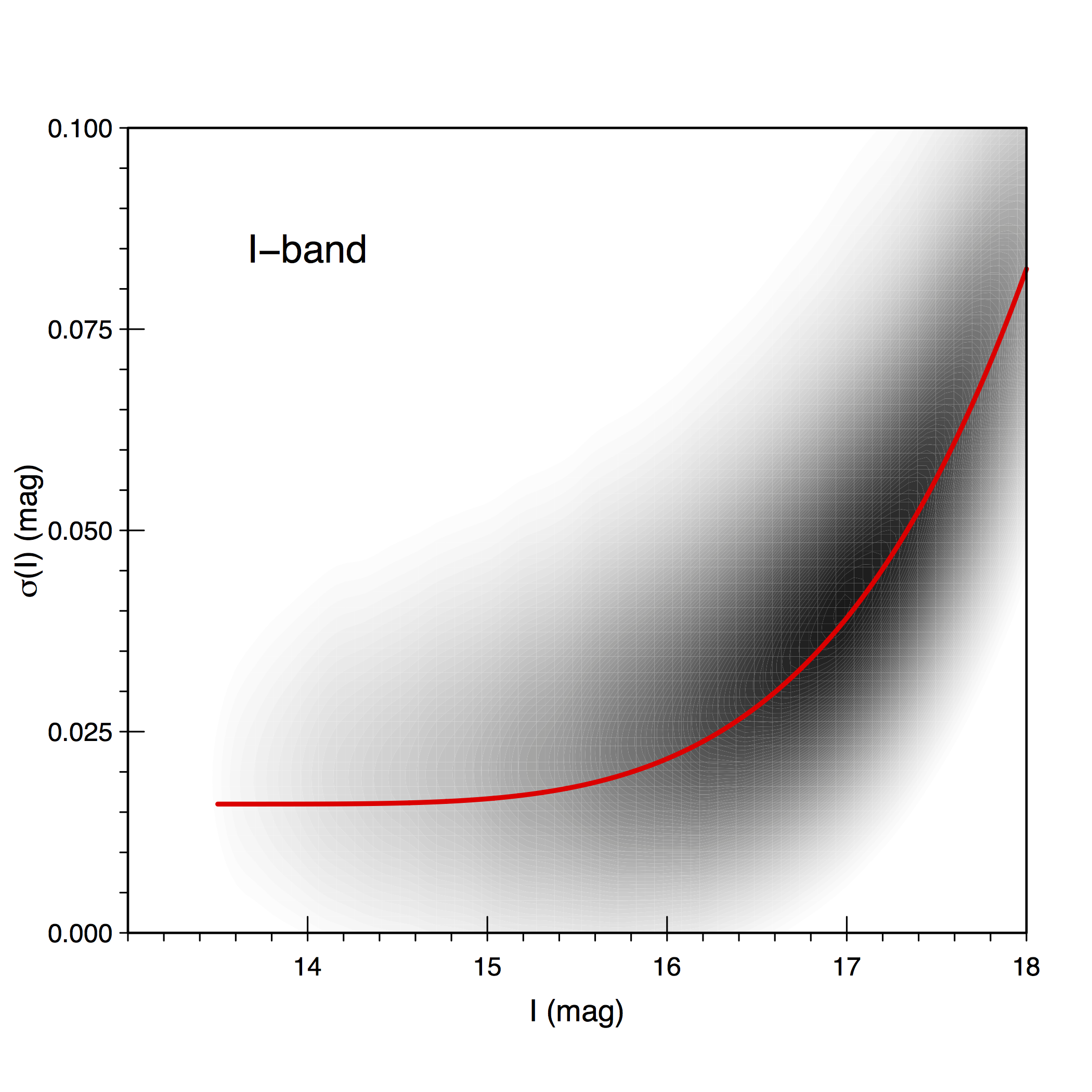}{0.31\textwidth}{(c)}
          }
    \caption{Scatter in the measured magnitudes for $\sim 1.7$ million stars in the CVSO data as a function of apparent magnitude. The plots are (a) $\sigma_{V}$ vs. V, (b) $\sigma_{R}$ vs. R and (c) $\sigma_{I}$ vs. I respectively. In grayscale we show the r.m.s.\ scatter versus mean apparent magnitude smoothed with a two-dimensional density kernel estimator \citep{venables02,Rmanual}. The colored solid lines are fits to the bulk of statistically non-variable stars, representing an empirical measure of the overall systemic photometric precision.
    \label{fig:sdvmag}}
\end{figure}

\begin{figure}[!ht]
    \centering
    \includegraphics[width=0.9\linewidth,clip,trim=0 50 0 50]{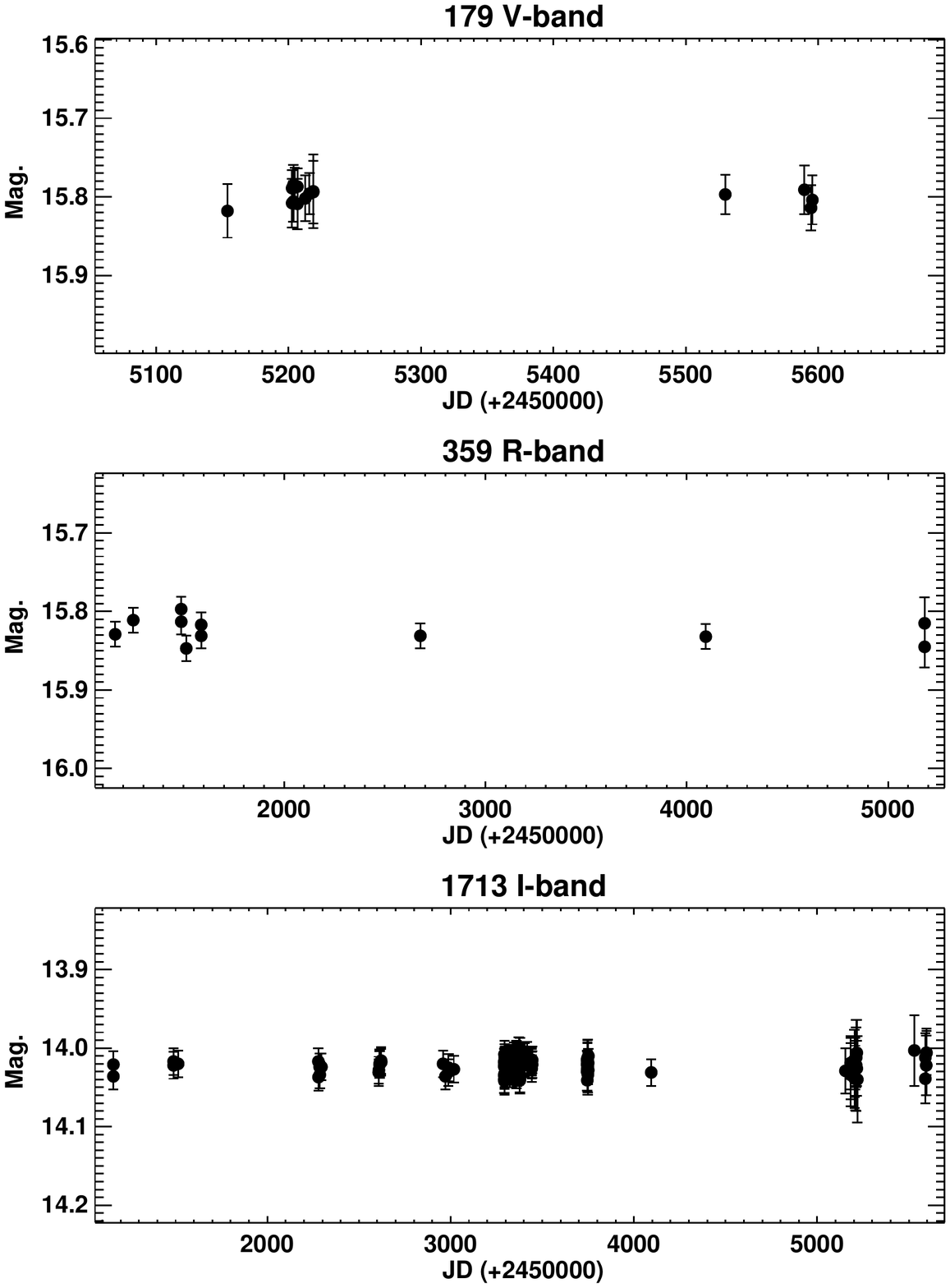}
    \caption{Example light curves of three non-variable stars from Fig.~\ref{fig:sdvmag}.}
    \label{fig:nonvar}
\end{figure}

To represent the noise properties of the light curves as a function of stellar apparent magnitude, we 
applied the following polynomial fits to the overall trends of light curve scatter versus magnitude for the full sample, as shown in Figure~\ref{fig:sdvmag} (and representative non-variable stars shown in Figure~\ref{fig:nonvar}):

\begin{gather}
\sigma_{V} = 2.7\times10^{-4}(<V> - 13.5)^{3.35} + 0.013\\
\sigma_{R} = 8.0\times10^{-5}(<R> - 13.5)^{4.1} + 0.013\\
\sigma_{I} = 1.2\times10^{-4}(<I> - 13.5)^{4.2} + 0.016
\end{gather}

\begin{figure}[!ht]
    \begin{minipage}[t]{0.4cm}
    \centering
    \rotatebox{90}{{\Large Period Recovered}}
    \end{minipage}%
    \begin{minipage}{\dimexpr\linewidth-2.20cm\relax}%
    \gridline{\fig{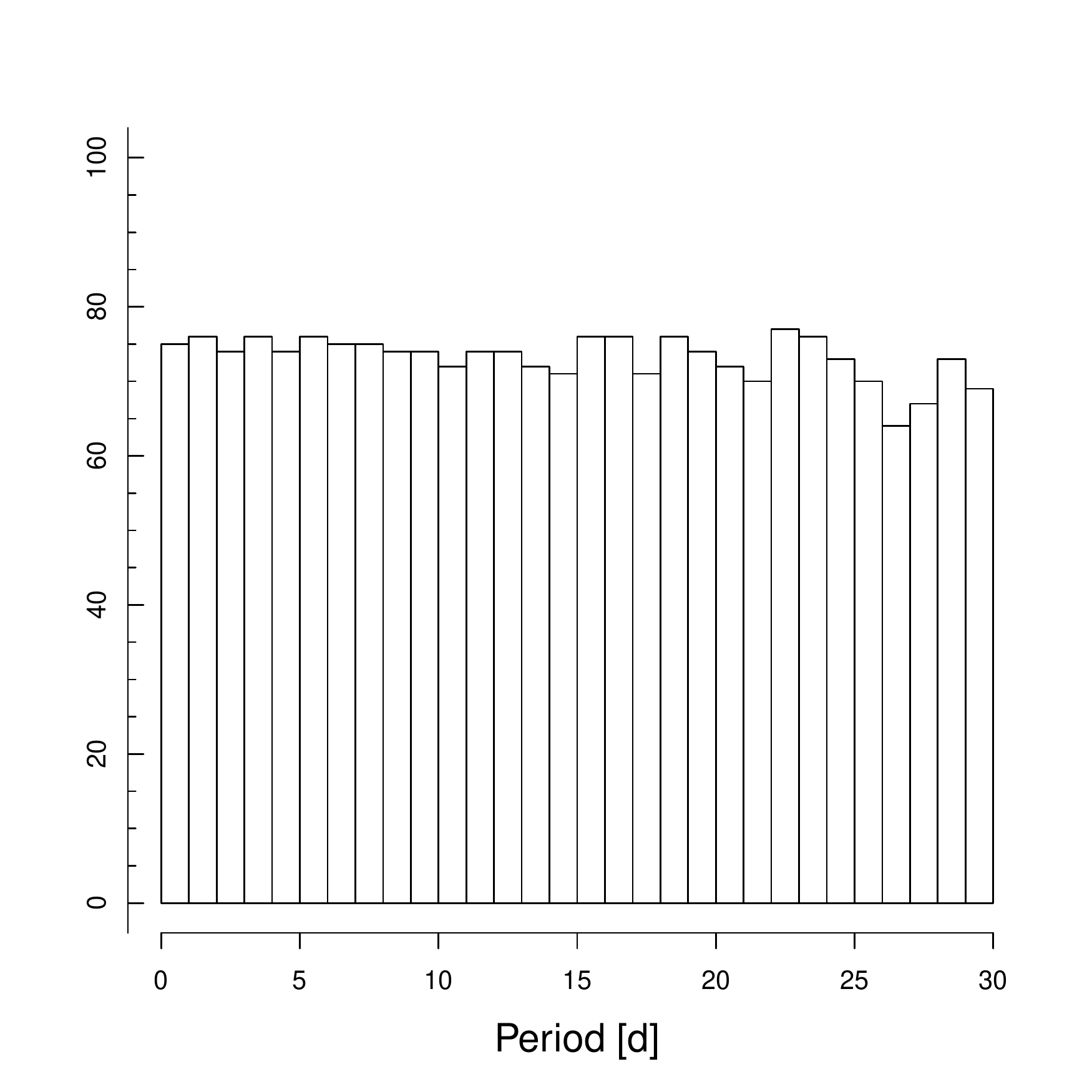}{0.45\textwidth}{(a)}
          \fig{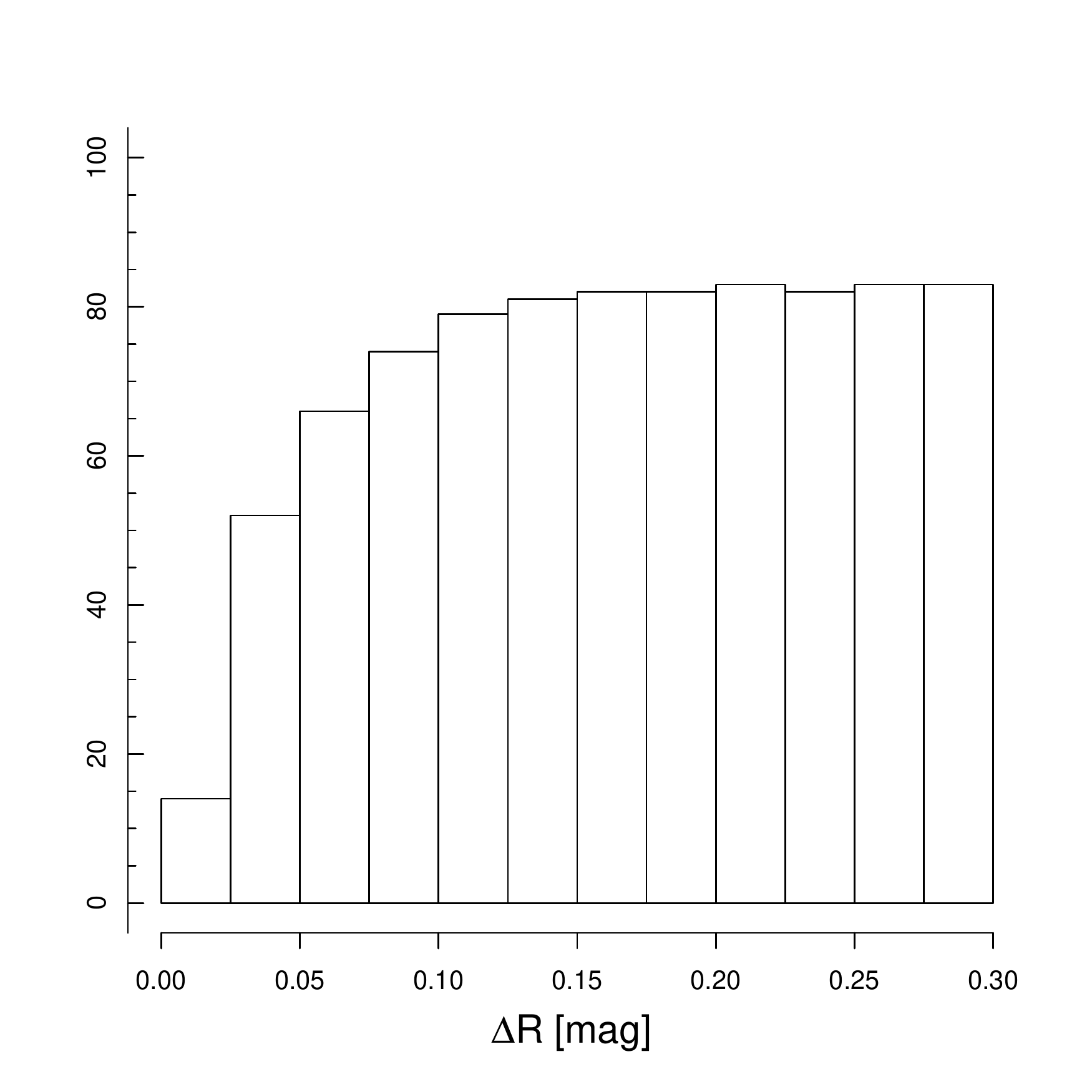}{0.45\textwidth}{(b)}}
    \gridline{
          \fig{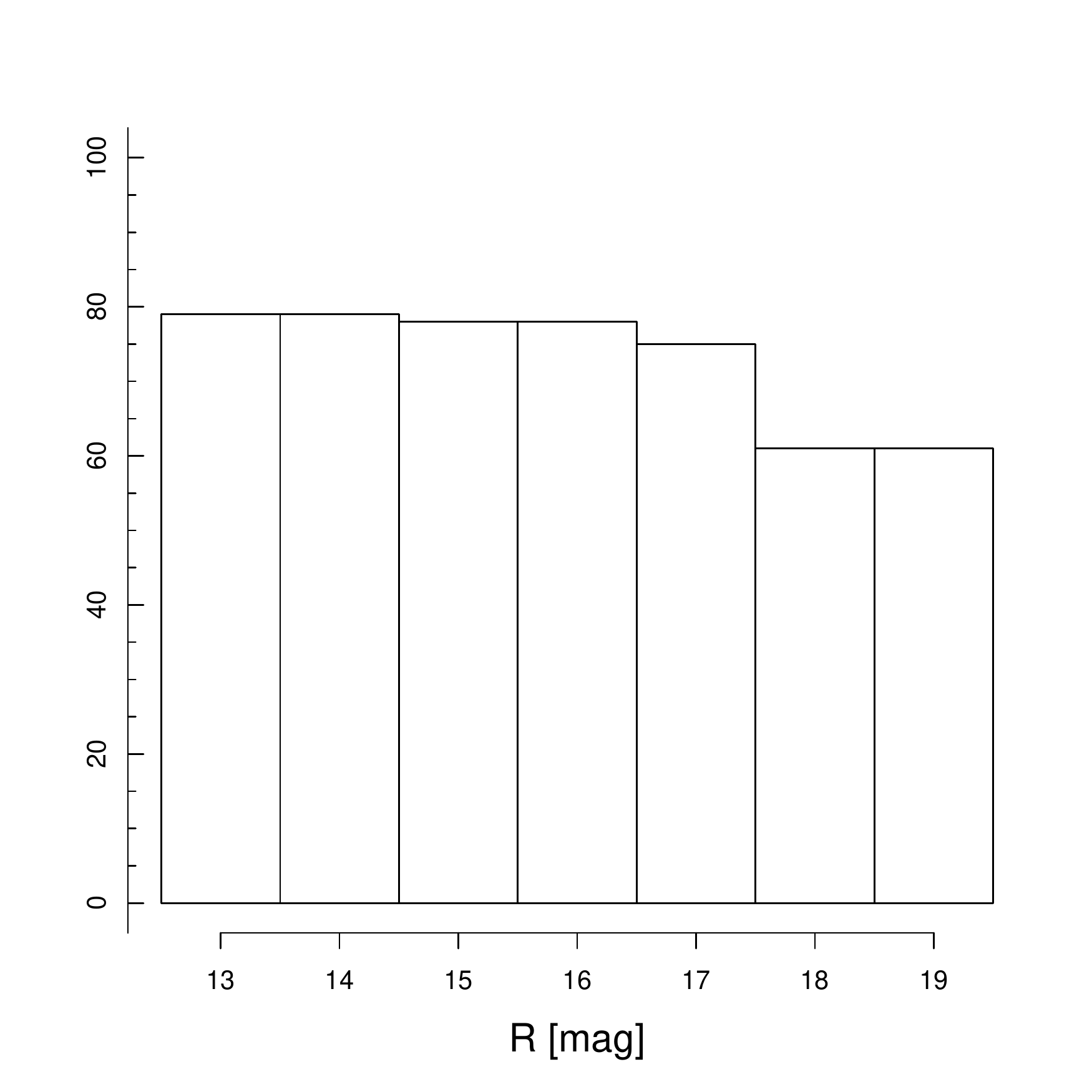}{0.5\textwidth}{(c)}
    }
    \caption{Period recoverability. (a) Period recoverability as a function of input periods, using synthetic light curves that have the same baseline, mean magnitude and amplitude of variability as the variable rotators. We injected sinusoids with periods between 0.1 and 30 days and amplitude 0.01 $\leq$ $\Delta R$ $\leq$ 0.3 mag into the light curves. A successful recovery means that the output period is within 10\% of the input period. (b) Same as (a), but showing period recoverability as a function of injected amplitude $\Delta R$. (c) Same as the other histograms, period recoverability as a function of R magnitude.
    \label{fig:vt}}
    \end{minipage}%
\end{figure}

Figure~\ref{fig:vt}(a) shows the result of period recoverability as a function of injected period, and it is seen that period detection sensitivity is roughly constant at 75\%, with a slight trend of decreasing sensitivity at the longest periods. 
Figures~\ref{fig:vt}(b) and (c) show period recoverability as a function of injected signal amplitude and of stellar apparent $R$ magnitude, respectively. 
Not surprisingly, period recoverability drops rapidly for low amplitude signals ($\lesssim$0.05 mag) and faint stars, where the photometric noise exceeds the amplitude of the rotational signal. 
Overall, these tests indicate that our data allow high recoverability (50--80\%) of rotation periods among stars in our sample with signals $\gtrsim$0.025 mag.

\section{Results:
Rotation Periods in Orion OB1 at \kgsins{4}--10 Myr}
\label{section:results}

We detected periods for a total of 564 PMS members of the Orion OB1 Association: 49 are Classical T Tauri stars (CTTSs), corresponding to $\sim$13\% of the total CTTS sample, 514 periodic weak-line T Tauri stars (WTTSs), corresponding to $\sim$32\% of the total WTTS sample, and one periodic G-type star with H$\alpha$ emission, which we classify as an Intermediate Mass TTS \citep[IMTTS;][]{calvet04}. Among the periodic TTSs, 540 out the 564 ($\sim 96$\%) can be assigned to either the OB1b or the OB1a subassociations as defined by \citep{briceno05}.  A total of 176 ($\sim$31\%) of the periodic TTSs are located in the $\sim 4$ Myr old OB1b, and the remaining 364 ($\sim 65$\%) are distributed across the older OB1a  subassociation, with 124 ($\sim 22$\%) concentrated in the $\sim 8$ Myr old  25 Ori cluster \citep{briceno07}. Only a small fraction ($\sim 4$\%) of the periodic stars are located in much younger regions like the surroundings of the Trapezium Cluster or projected against the Orion B cloud (Figure \ref{fig:sd}).

\begin{figure}[!ht]
\centering
\includegraphics[width=0.5\textwidth]{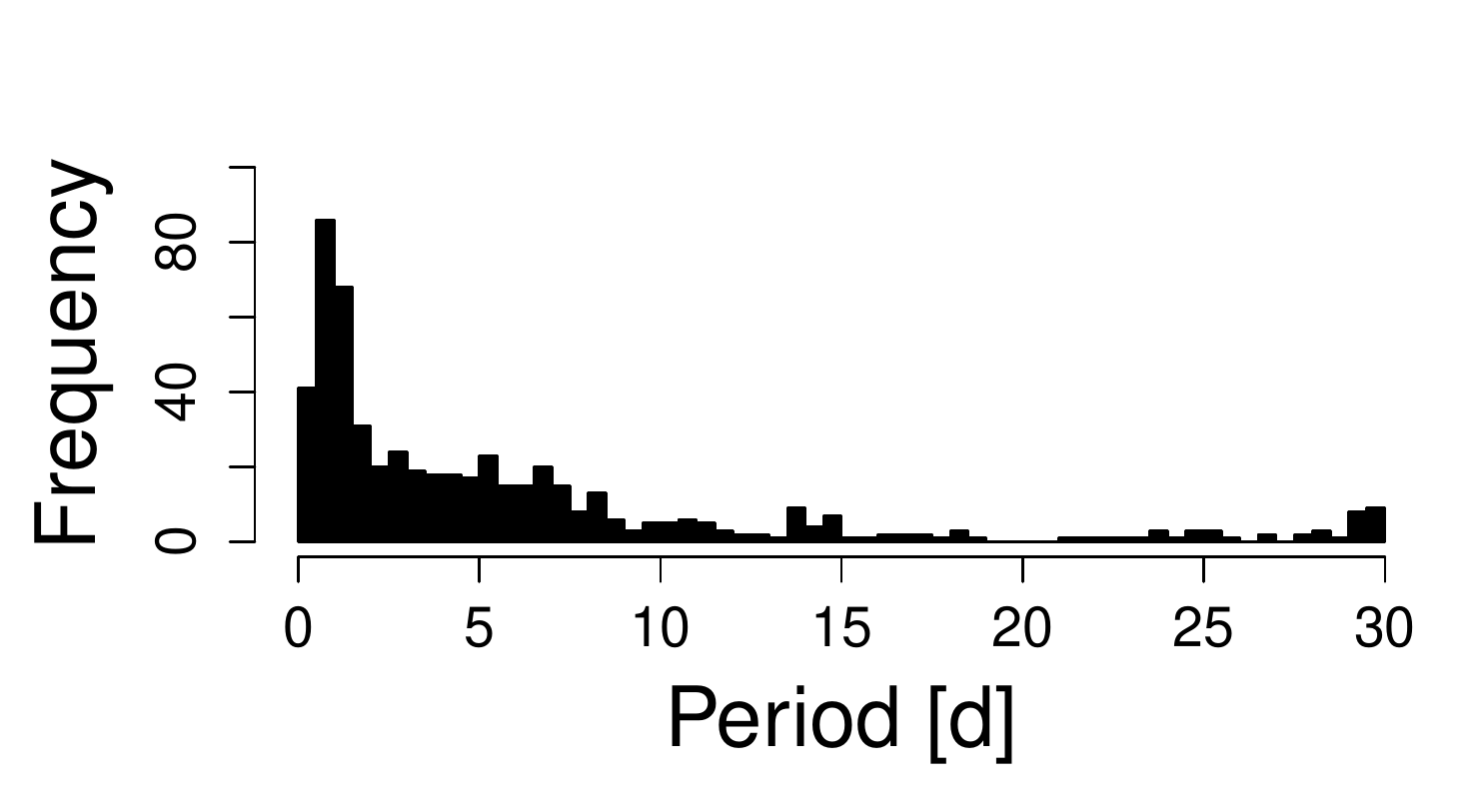}
\caption{Rotation period distribution of all periodic TTSs \kgsins{in this study. Note that periods within $\pm$0.05~d of integer days are excluded as aliases from each 1~d period bin (see the text)}.}
\label{fig:total}
\end{figure}

\begin{figure}
    \gridline{\fig{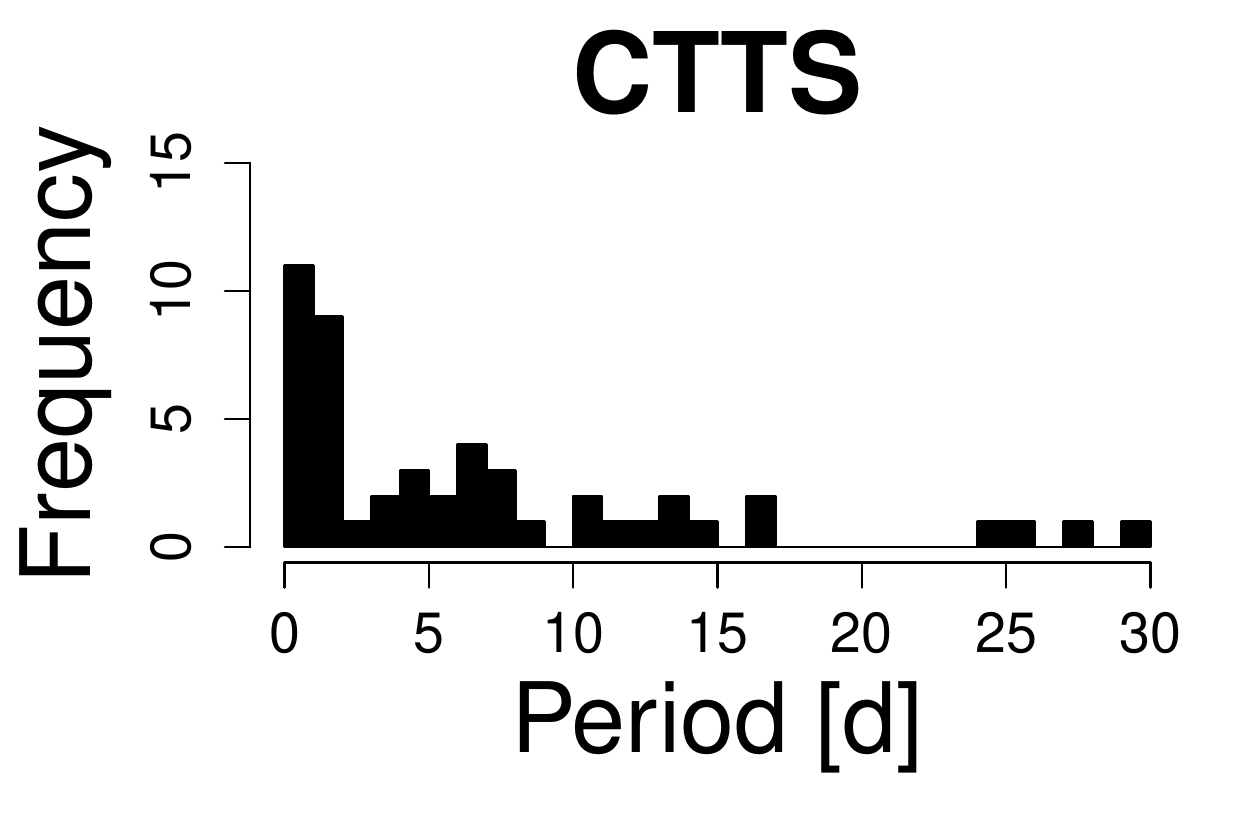}{0.45\textwidth}{(a)}
          \fig{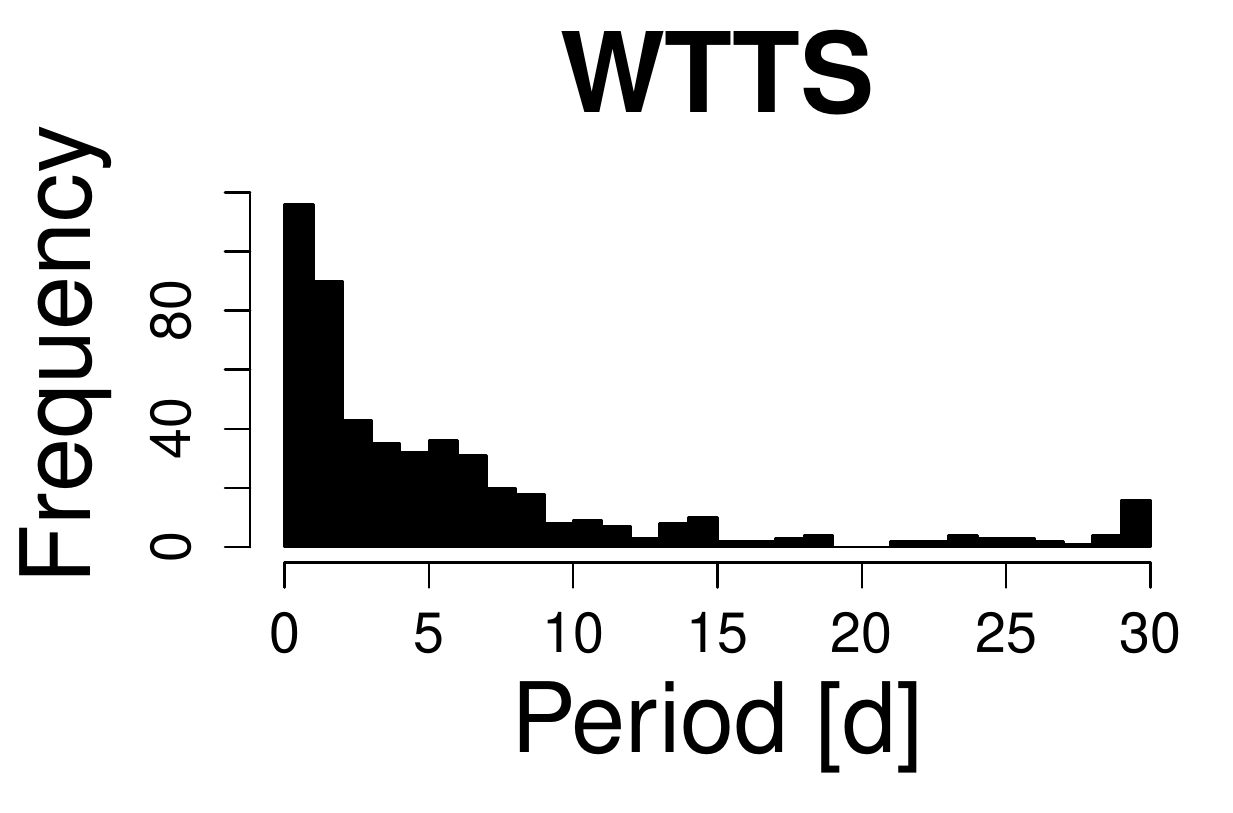}{0.45\textwidth}{(b)}
    }
    \caption{Rotation period distribution by star type. \kgsins{Note that periods within $\pm$0.05~d of integer days are excluded as aliases from each 1~d period bin (see the text).} (a) The rotation period distribution of CTTSs with mean rotation period 6.87~d. (b) The rotation period distribution of WTTSs with the mean rotation period 5.84~d. The Anderson-Darling test of these distributions has a p-value of 0.44, which implies that there is no significant difference between the rotation periods of CTTS and WTTS. \label{fig:rot_type}}
\end{figure}

The overall distribution of rotation periods for the 564 periodic stars in our sample is shown in Figure~\ref{fig:total}, and in
Figure~\ref{fig:rot_type} the distribution is divided into that for CTTSs and that for WTTSs. The two distributions appear broadly very similar, with mean rotation periods for CTTSs and WTTSs of 6.87~d and 5.84~d, respectively. To quantify this, we used a K-Sample Anderson-Darling (A-D) test \citep{scholz87} as implemented in the \textit{kSamples} package \citep{kSamples} in {\tt R} to test whether the two samples are significantly different. The A-D test is a non-parametric hypothesis test that measures the likelihood of two univariate distributions being drawn from the same parent population. 
We used the critical value $\alpha = 0.05$ for this test.
The A-D test $p$-value in this case is 0.44, which implies that the CTTS and WTTS rotation period distributions are indeed consistent with having been drawn from the same population (Figure \ref{fig:total}). 

\begin{figure}
    \gridline{\fig{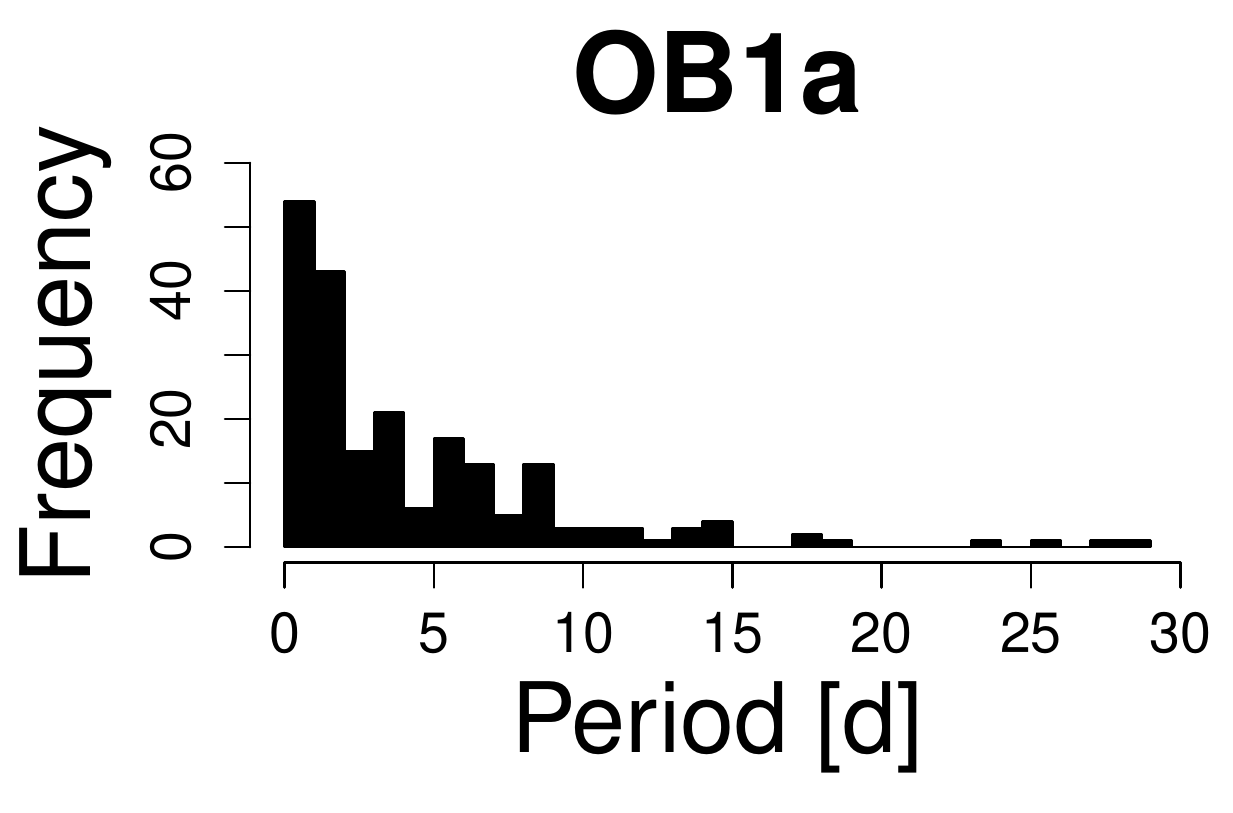}{0.45\textwidth}{(a)}
          \fig{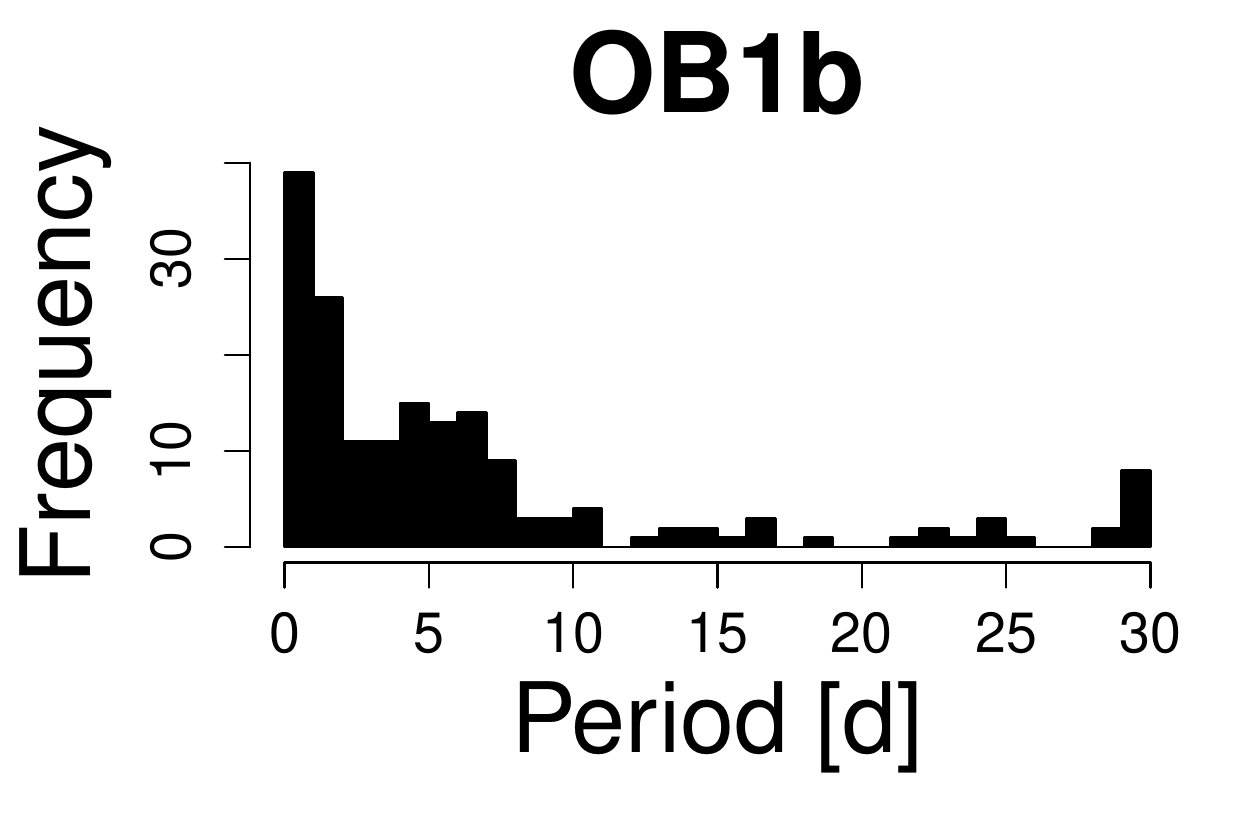}{0.45\textwidth}{(b)}
          }
    \gridline{\fig{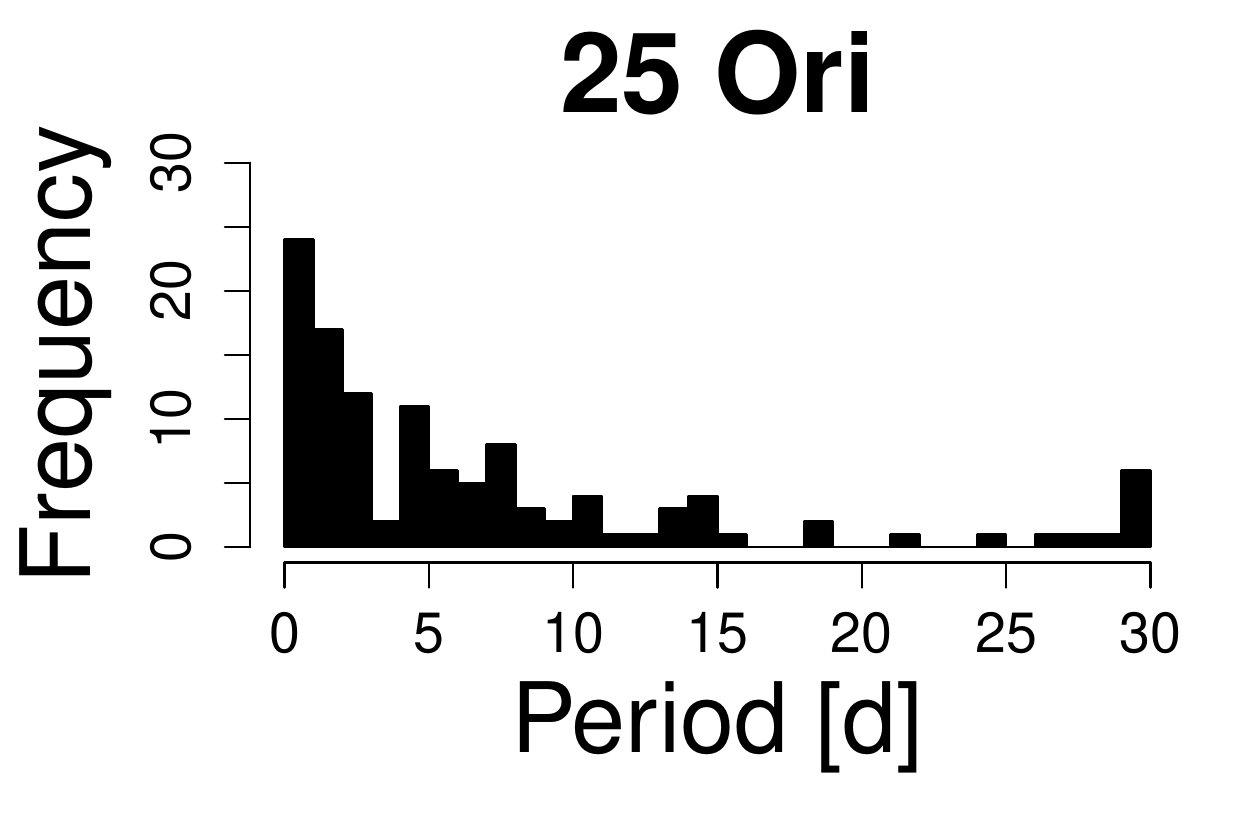}{0.5\textwidth}{(c)}
          }
    \caption{Rotation period distribution by star location. \kgsins{Note that periods within $\pm$0.05~d of integer days are excluded as aliases from each 1~d period bin (see the text).} (a) The rotation period distribution of TTS in OB1a with mean rotation period 4.37~d. TTS in OB1a are $\sim$ 10 Myr old. (b) The rotation period distribution of TTS in OB1b with mean rotation period 6.55~d. TTS in OB1b are $\sim$ 4 Myr old. The Anderson-Darling test of these distributions has a p-value of 0.025, which implies that at 97.5\% confidence, older TTS rotate faster than  younger TTS. (c) The rotation period distribution of TTS in 25 Ori with mean rotation period 6.98~d.   \label{fig:rot_loc}}
\end{figure}

Next, we assess the evidence for possible changes in rotational properties as a function of stellar age. Figure~\ref{fig:rot_loc} shows the rotation period distributions based on Orion OB1 sub-region. 
The mean rotation periods for OB1b ($\sim$4 Myr) and OB1a($\sim$10 Myr) are 6.55~d and 4.37~d, respectively, and the A-D test $p$-value for these distributions is 0.025, implying that there is indeed a difference in the distributions between 4 and 10 Myr at 97.5\% confidence. 
That there is a difference in rotation periods among stars in different aged clusters but not between CTTSs and WTTSs as a whole is consistent with previous work that finds significant age overlaps between CTTSs and WTTSs. Thus, it is by examining stars according to their cluster ages that we detect the effects of rotational evolution with age. 
In particular, 
we can infer from this result that stars at 10 Myr rotate faster than their counterparts at 4 Myr, 
suggesting that PMS stars experience spinup as they contract during this phase of evolution when protoplanetary disks have mostly or entirely dissipated.

\section{Discussion: Testing Rotational Evolution Models From 1 Myr to 10 Myr}
\label{sec:disc}

\begin{figure}[!ht]
\centering
\includegraphics[width = 0.75\textwidth]{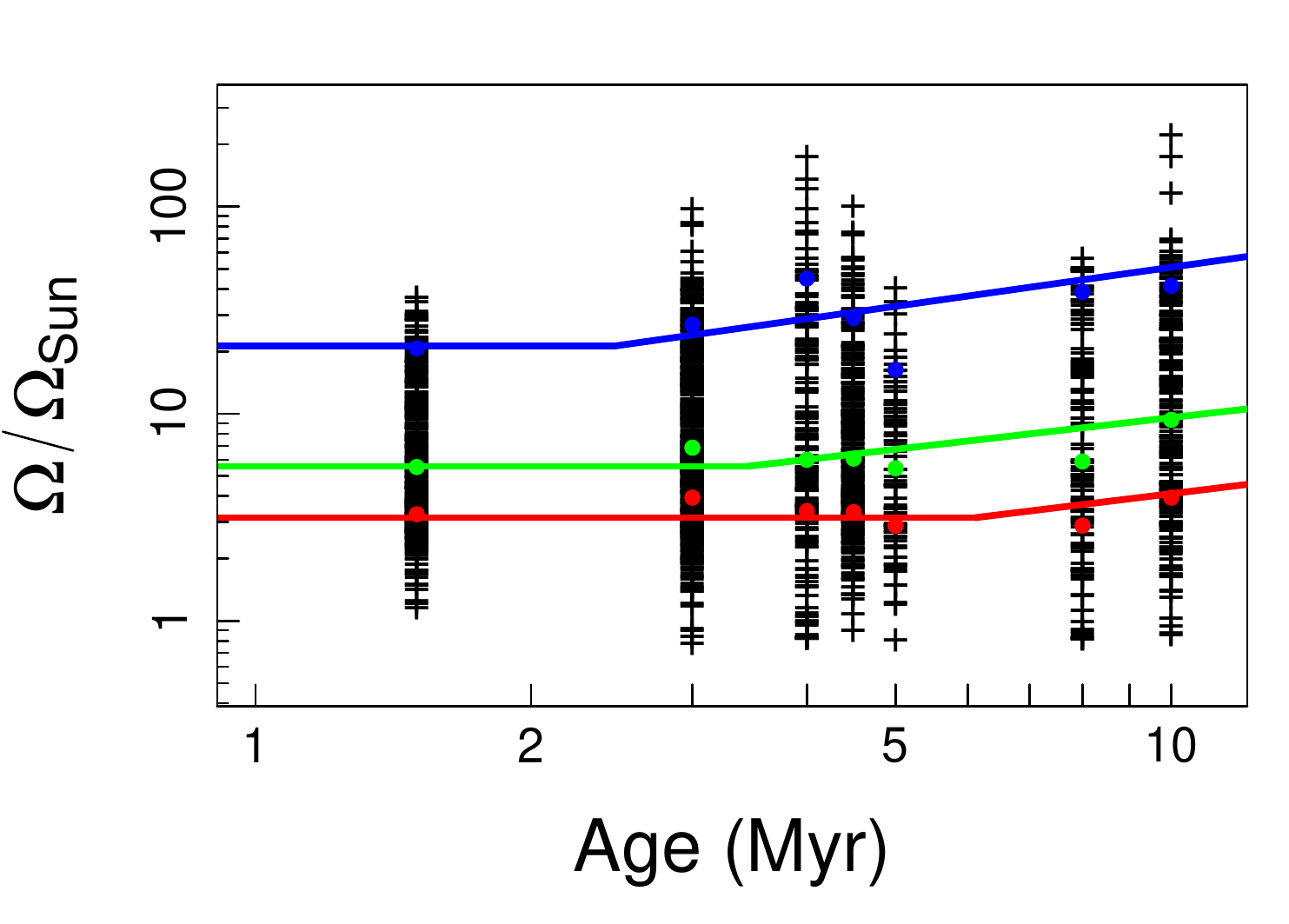}
\caption{Distributions of stellar angular velocities (normalized to the Sun) as a function of stellar age, for stars in the 4--10 Myr old Orion OB1a sub-regions from this study together with stars in various 1--5 Myr old regions from the literature (see Table~\ref{tab:comp}). The red, green, and blue circles at each age represent the 25$^{th}$, 50$^{th}$, and 90$^{th}$ percentile of angular velocities, respectively. Here, $\Omega_\odot = 0.26$ rad~d$^{-1}$ \citep{snodgrass90}. 
The solid curves represent the models proposed by \citet{gallet15} to describe the evolution of surface angular velocity for 0.5 $M_{\odot}$ stars.}
\label{fig:comp}
\end{figure}

We can also compare our resulting rotation period distributions with that of the much younger stars ($\sim$1--2 Myr) in the Orion Nebula Cluster \citep[e.g.,][]{stassun99}. 
The rotation period distribution of TTSs in the ONC 
has a mean of 5.15~d. 
We again applied the A-D test for ONC vs.\ OB1a and for ONC vs.\ OB1b, 
finding $p$-values 
$<$10$^{-6}$, implying that the rotation periods in the ONC and OB1a/OB1b regions are distinct. As with the change in rotation properties over the 4--10 Myr age range, this implies that 
stars at $\sim$10 Myr rotate faster
on average than at $\sim$1--2 Myr.
As can be seen in Figure~\ref{fig:comp}, the models predict constant angular velocity during first few Myr for all stars, followed by a linear increase in the angular velocity resulting from conservation of angular momentum after the disk-related braking phenomena cease. In these models, the fastest rotators begin this spinup at 2.5 Myr whereas the slowest rotators do not begin the spinup until an age of 6 Myr.

In order to paint a more complete picture of the rotational evolution of young, low-mass stars, we combine our resulting rotation period distributions for Orion OB1 stars at 4--10 Myr with previously published results for star-forming regions at younger ages (1--5 Myr), as summarized in Table~\ref{tab:comp}. We then compare these observed distributions to
rotational evolution models \citep{gallet15}, as shown in Figure~\ref{fig:comp}. 
For simplicity we adopt the model predictions for stars of 0.5 M$_\odot$ because the median stellar masses of the seven clusters are all approximately 0.5 M$_\odot$ (see Table~\ref{tab:comp} and see Section \ref{section:data}).
These models attempt to explain the evolution of the surface rotation of young, low-mass stars through a combination of stellar core-envelope decoupling, conservation of angular momentum during the stellar contraction, and angular momentum removal via accretion-related processes during the first few Myr. 

\begin{deluxetable}{cccccccc}


\tablecaption{Angular velocity distributions of the stellar clusters considered in this study. Here, $\Omega_\odot = 0.26$ rad~d$^{-1}$ \citep{snodgrass90}.  Nominal cluster ages are taken from \citet{briceno05,briceno07,Henderson2012} and references therein. \label{tab:comp}}

\tablehead{\colhead{Cluster Name} & \colhead{Age} & \colhead{Mean} & \colhead{Ref} & \colhead{$\Omega_{Mean}/\Omega_\odot$} & \colhead{$\Omega_{25}/\Omega_\odot$} & \colhead{$\Omega_{50}/\Omega_\odot$} & \colhead{$\Omega_{90}/\Omega_\odot$} \\ 
\colhead{(---)} & \colhead{(Myr)} & \colhead{(Mass)} & \colhead{(---)} & \colhead{(---)} & \colhead{(---)} & \colhead{(---)} & \colhead{(---)} } 

\startdata
ONC & 1.5 & 0.53 & 1 & 4.75 & 3.27 & 5.53 & 20.64 \\
NGC 2264 & 3 & 0.48 & 2 & 5.29 & 3.93 & 6.85 & 26.91 \\
OB1b4 & $\sim$0.5 & 4 & 3.74 & 3.39 & 6.00 & 44.90 \\
NGC 2362 & 4.5 & 0.61 & 3 & 4.80 & 3.35 & 6.08 & 29.01 \\
IC 348 & 5 & 0.34 & 5 & 4.05 & 2.88 & 5.45 & 16.32 \\
25 Ori & 8 & $\sim$0.5 & 4 & 3.51 & 2.88 & 5.88 & 38.71 \\
OB1a & 10 & $\sim$0.5 & 4 & 5.60 & 3.93 & 9.35 & 41.48 \\
\enddata

\end{deluxetable}


\kgsins{Importantly,} the rotation periods newly provided in this study, and in particular those for 25 Ori ($\sim$8 Myr) and OB1a ($\sim$10 Myr) extend previous observational constraints for the first time into the age regime of $\gtrsim$5 Myr where the models predict the critical transition from constant to increasing angular velocity.

There is a hint from Fig.~\ref{fig:comp} that the mean period for 25 Ori is more akin to that of the 4--5 Myr regions than to that of the 10 Myr old region, despite the nominal age of 25 Ori being 8 Myr \citep{briceno07}. It may be that the age of 25 Ori should be revised younger by $\sim$1--1.5 Myr, a question that is explored further in Brice\~no et al.\ (2016, in preparation). On the other hand, the much younger ($\sim 3$ Myr old) cluster NGC 2264
shows a mean angular rotation which is more comparable to the older OB1a (see Table \ref{tab:comp}). These apparent inconsistencies show the large scatter
in rotation properties among stars in a single (supposedly coeval) population,
which could be the result of a wide range of initial conditions, like the
primordial disk fraction \citep[see][]{muzerolle_flaherty2008,ingleby2012}.

Nonetheless, the slightly different mean rotation period for 25 Ori does not alter our basic conclusion that there is overall a very good agreement between the new data presented here and the model predictions in Fig.~\ref{fig:comp}, suggesting that indeed low-mass PMS stars in Orion OB1 are undergoing the transition to angular momentum conservation as predicted.

\section{Summary and Conclusions}
\label{section:conclusion}
We have measured rotation periods for 564 T Tauri stars in the Orion OB1 star-forming complex.
\kgsins{As such, this study provides one of the largest samples of TTS rotation periods in the literature, particularly among the previously poorly sampled age range of 4--10 Myr.}
Among the periodic stars, 49 are CTTS, 514 are WTTS, and one is classified as an IMTTS of Ge type; this corresponds to $\sim$13\% CTTS and $\sim$32\% WTTS being periodic among the complete sample of 1974 objects. As expected from the CVSO being a survey targeted at the somewhat older, off-cloud populations of Orion OB1, the vast majority ($\sim 96$\%) of the periodic stars are located in the $\sim 4$ Myr old OB1b subassociation (176 stars) and the $\sim 10$ Myr old OB1a subassociation (340 stars), with 124 of the 340 concentrrated in the $\sim 8$ Myr old 25 Ori cluster; only 24 stars are projected on the younger ($\sim 1-2$ Myr ) regions like the area around the Trapezium cluster or on the Orion B cloud. 

Therefore, our sample mostly spans the little studied $\sim 4-10$ Myr age range.
At these ages, we do not detect significant differences between the rotation period distributions of CTTS vs.\ WTTS. 

However, we do find \kgsdel{strong} observational evidence of rotational evolution from 4 to 10 Myr; the rotation periods spin up as the stars contract,
consistent with conservation of angular momentum. 
Indeed, the rotation period distributions newly reported here at ages of 5--10 Myr fill in an important gap in previously available rotation period data for low-mass PMS stars and permit a direct test of critical predictions of PMS angular momentum evolution models. We find that the observations are highly consistent with the model predictions. Evidently, low-mass PMS stars transition to a phase of angular momentum conservation at an age of 5--10 Myr, coinciding with the time when protoplanetary disks become mostly or entirely dissipated.


\acknowledgements
\cbins{This research is based partly on observations collected at the J\"urgen Stock 1m Schmidt telescope 
of the National Observatory of Llano del Hato Venezuela, operated by CIDA
for the Ministerio del Poder Popular para la Ciencia y Tecnolog{\'\i}a, Venezuela,
and on observations obtained at Kitt Peak National Observatory, National Optical Astronomy Observatory (NOAO Prop. ID: 2005B-0529; PI: K. Stassun), 
which is operated by the Association of Universities for Research in Astronomy (AURA) under cooperative agreement with the National Science Foundation. 
This study was conducted as part of the the Cerro Tololo Inter-American Observatory REU Program and the Vanderbilt University REU Program, which are supported by the National Science Foundation under grants AST-1062976 and PHY-1263045 respectively. We thank Herbert Pablo for assistance with the Kitt Peak observations.} 

\bibliographystyle{apj} 
\bibliography{main-v2}

\clearpage

\appendix
\section{Appendix: Light Curves}
\label{section:applc}

\section{Appendix: Wavelet Transform}
\label{section:appwt}

\section{Appendix: Period Table}
\label{section:apptable}
\end{document}